\documentclass[twoside,prl,letterpaper,preprintnumbers,superscriptaddress,notitlepage,onecolumn]{revtex4}

\usepackage{mathptmx}
\usepackage{helvet}
\usepackage{graphicx,fancyhdr}
\usepackage{dashbox,color}%
\usepackage{bm}
\usepackage{sidecap}%
\definecolor{lightblue}{rgb}{0.6,0.9,1}
\definecolor{myrefblue}{rgb}{0.1,0.6,1}
\definecolor{myblue}{rgb}{0,0,0}
\definecolor{nmat}{rgb}{0.7,0.04,0.26}

\newcommand{\sfcaopt}{\mbox{SmFe$_{\mathsf{0.92}}$Co$_{\mathsf{0.08}}$AsO}}
\newcommand{\nfaofopt}{\mbox{NdFeAsO$_{\mathsf{0.6}}$F$_{\mathsf{0.4}}$}}

\begin{document}
\pagestyle{fancy}

\renewcommand{\headrule}{\vskip-3pt\hrule width\headwidth height\headrulewidth \vskip-\headrulewidth}

\fancypagestyle{plainfancy}{%
\lhead{}
\rhead{}
\chead{}
\lfoot{}
\cfoot{}
\rfoot{\bf\scriptsize\textsf{\thepage}}
\renewcommand{\headrulewidth}{0pt}
\renewcommand{\footrulewidth}{0pt}
}

\fancyhead[LE,RO]{}
\fancyhead[LO,RE]{}
\fancyhead[C]{}
\fancyfoot[LO,RE]{}
\fancyfoot[C]{}
\fancyfoot[LE,RO]{\bf\scriptsize\textsf{\thepage}}

\renewcommand\bibsection{\section*{\sffamily\bfseries\normalsize {References}\vspace{-10pt}\hfill~}}
\newcommand{\mysection}[1]{\section*{\sffamily\bfseries\normalsize {#1}\vspace{-10pt}\hfill~}}
\renewcommand{\subsection}[1]{\noindent{\bfseries\normalsize #1}}
\renewcommand{\bibfont}{\fontfamily{ptm}\footnotesize\selectfont}
\renewcommand{\figurename}{Fig.}
\renewcommand{\refname}{References}
\renewcommand{\bibnumfmt}[1]{#1.}

\makeatletter
\long\def\@makecaption#1#2{%
  \par
  \vskip\abovecaptionskip
  \begingroup
   \small\rmfamily
   \sbox\@tempboxa{%
    \let\\\heading@cr
    \textbf{#1.\hskip1pt} #2%
   }%
   \@ifdim{\wd\@tempboxa >\hsize}{%
    \begingroup
     \samepage
     \flushing
     \let\footnote\@footnotemark@gobble
     \textbf{#1.\hskip1pt} #2\par
    \endgroup
   }{%
     \global \@minipagefalse
     \hb@xt@\hsize{\hfil\unhbox\@tempboxa\hfil}%
   }%
  \endgroup
  \vskip\belowcaptionskip
}%
\makeatother

\thispagestyle{plainfancy}

\fontfamily{helvet}\fontseries{bf}\selectfont
\mathversion{bold}
\begin{widetext}
\begin{figure}
\vskip0pt\noindent\hskip-0pt
\hrule width\headwidth height\headrulewidth \vskip-\headrulewidth
\hbox{}\vspace{4pt}
\hbox{}\noindent\vskip10pt\hbox{\noindent\huge\sffamily\textbf{High-temperature superconductivity from fine-tuning}}\vskip0.05in\hbox{\noindent\huge\sffamily\textbf{of Fermi-surface singularities in iron oxypnictides}}
\vskip10pt
\hbox{}\noindent\begin{minipage}{\textwidth}\flushleft
\renewcommand{\baselinestretch}{1.2}
\hskip-10pt\large\sffamily A.~Charnukha$^{1,2,\dagger}$, D.~V.~Evtushinsky$^{1}$, C.~E.~Matt$^{3,4}$, N.~Xu$^{3,5}$, M.~Shi$^{3}$, B.~B\"uchner$^{1}$, N.~D.~Zhigadlo$^{4}$, B.~Batlogg$^{4}$ \&~S.~V.~Borisenko$^{1}$
\end{minipage}
\end{figure}
\end{widetext}

\begin{figure}[!h]
\begin{flushleft}
{\footnotesize\sffamily
$^1$Leibniz Institute for Solid State and Materials Research, IFW, D-01069 Dresden, Germany, $^2$Physics Department, University of California--San Diego, La Jolla, CA 92093, USA, $^3$Swiss Light Source, Paul Scherrer Institut, CH-5232 Villigen PSI, Switzerland, $^4$Laboratory for Solid State Physics, ETH Zurich, CH-8093 Zurich, Switzerland, $^5$Institute of Condensed Matter Physics, \'Ecole Polytechnique F\'ed\'erale de Lausanne,
CH-1015 Lausanne, Switzerland. $^\dagger$ E-mail: acharnukha@ucsd.edu.}
\end{flushleft}
\end{figure}

\fontsize{8pt}{8pt}\selectfont
\renewcommand{\baselinestretch}{0.9}
\noindent\sffamily\bfseries{In the family of the iron-based superconductors, the $RE$FeAsO-type compounds (with $RE$ being a rare-earth metal) exhibit the highest bulk superconducting transition temperatures ($T_{\mathrm{c}}$) up to $55\ \textrm{K}$ and thus hold the key to the elusive pairing mechanism. Recently, it has been demonstrated that the intrinsic electronic structure of \sfcaopt\ ($T_{\mathrm{c}}=18\ \textrm{K}$) is highly nontrivial and consists of multiple band-edge singularities in close proximity to the Fermi level. However, it remains unclear whether these singularities are generic to the $RE$FeAsO-type materials and if so, whether their exact topology is responsible for the aforementioned record $T_{\mathrm{c}}$. In this work, we use angle-resolved photoemission spectroscopy (ARPES) to investigate the inherent electronic structure of the \nfaofopt\ compound with a twice higher $T_{\mathrm{c}}=38\ \textrm{K}$. We find a similarly singular Fermi surface and further demonstrate that the dramatic enhancement of superconductivity in this compound correlates closely with the fine-tuning of one of the band-edge singularities to within a fraction of the superconducting energy gap $\Delta$ below the Fermi level. Our results provide compelling evidence that the band-structure singularities near the Fermi level in the iron-based superconductors must be explicitly accounted for in any attempt to understand the mechanism of superconducting pairing in these materials.}
\mathversion{normal}
\normalfont\normalsize

After seven years of intensive research, the mechanism of superconductivity in the iron-based superconductors remains unknown. The observation of the Fermi-surface topology drastically different from the early predictions of the {\it ab initio} calculations in some compounds~\cite{2009PhyC469614M,Borisenko_BKFA_FS_NormState2009,PhysRevB.79.054517,Charnukha_SFCAO_2015} has cast doubt on the theoretical description of superconductivity based on this simplified picture and necessitates the adoption of the true Fermi surface, a paradigm shift that is now beginning to occur~\cite{PhysRevB.88.174516,PhysRevB.90.035104}. In this view, the identification of essential electronic features bearing directly on the superconducting pairing mechanism is of vital importance. Recent observation of a highly nontrivial electronic structure in \sfcaopt\ shaped by multiple band-edge singularities in the vicinity of the Fermi level ($E_\mathrm{F}$) implies, should it prove to be generic for the 1111-type compounds, that very small changes on the order of tens of meV can potentially lead to drastic changes in the Fermi-surface topology and strongly affect the superconducting state. We employ ARPES to show that the bulk electronic structure of \nfaofopt\ is, at first sight, virtually identical to that of sister \sfcaopt\ with a twice lower $T_{\mathrm{c}}$, with one essential difference: the top of the hole band located at $\approx6\ \textrm{meV}$ above the Fermi level in \sfcaopt\ (Ref.~\onlinecite{Charnukha_SFCAO_2015}) sinks to $2.3\ \textrm{meV}$ {\it below} the Fermi level, entirely eliminating the Fermi surface at the center of the Brillouin zone. And yet this band remains within the reach of superconducting pairing interaction and, intriguingly, hosts a superconducting energy gap of $10.5\ \textrm{meV}$ --- almost 10 times larger than that in \sfcaopt\ (Ref.~\onlinecite{Charnukha_SFCAO_2015}). This observation suggests that the ``fine-tuning'' of the aforementioned hole band dramatically enhances superconductivity in \nfaofopt.

\begin{figure}[!b]
\vskip40pt
\end{figure}

\begin{figure*}[!t]
\includegraphics[width=0.8\textwidth]{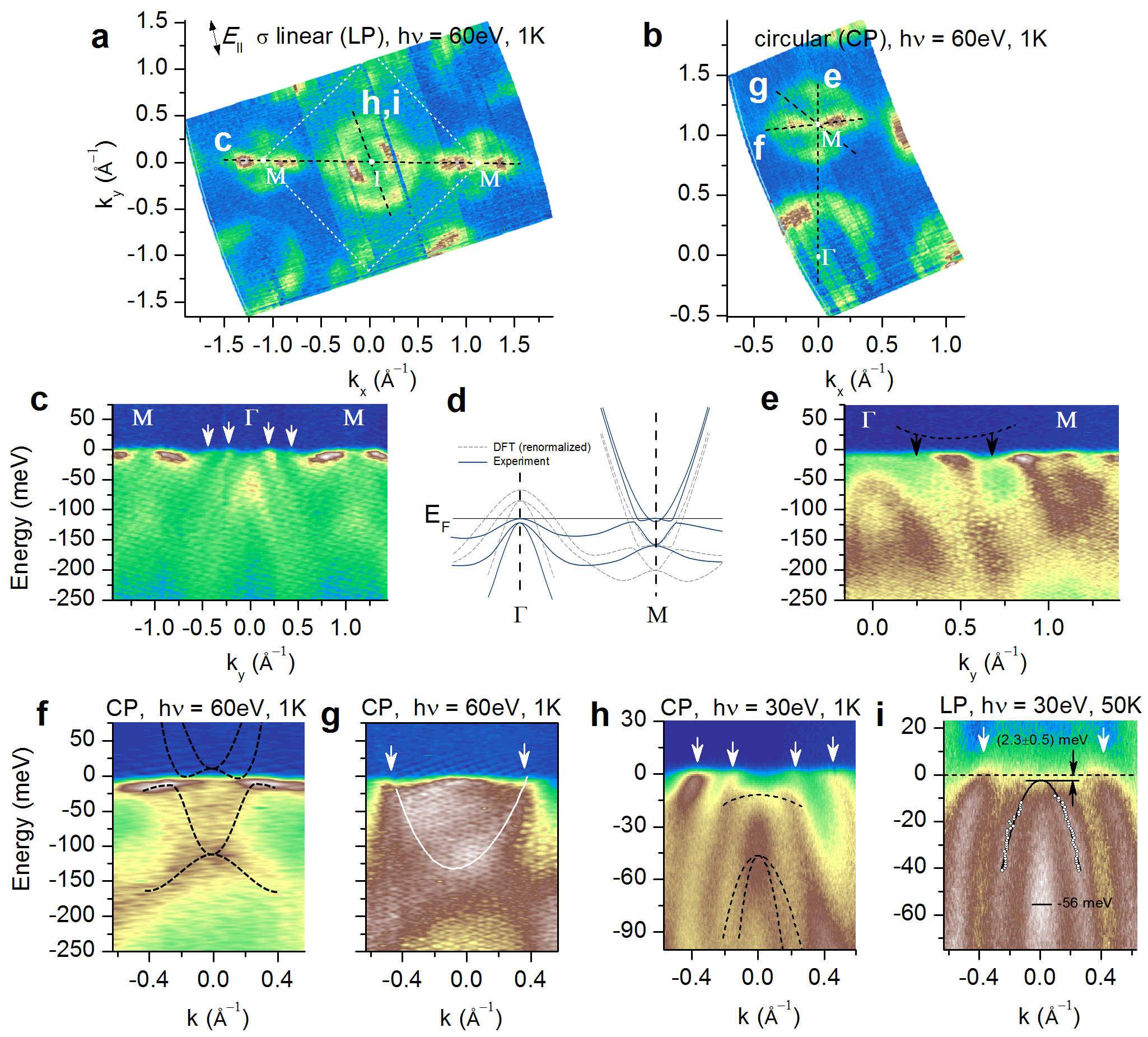}
\caption{\label{fig:fs}
~\textbf{a,b,} Constant-energy maps obtained by integrating the photoemission intensity within $15\ \textrm{meV}$ around $E_\mathrm{F}$, recorded at $T=1\ \textrm{K}$ using $60\ \textrm{eV}$ photons with linear horizontal (\textbf{a}) and circular (\textbf{b}) polarization. White dotted square shows the {2-Fe} Brillouin zone.~\textbf{c,} Energy-momentum cut (EMC) along the line shown in~\textbf{a}. White arrows in panels~\textbf{c,g}--\textbf{i} indicate the surface-related bands giving rise to the large circular features around $\Gamma$ and $M$ in~\textbf{a,b}.~\textbf{d,} Schematic illustration of the effect of high-energy electronic correlations on the {\it ab initio} band structure, as reported in ref.~\onlinecite{Charnukha_SFCAO_2015}.~\textbf{e}--\textbf{i,} EMCs along the lines shown in~\textbf{a,b}.~\textbf{e,} The intensity distribution near $\Gamma$ shows the bulk electronic structure more clearly: a hole band terminating near $E_\mathrm{F}$ together with another broad holelike dispersive feature at $56\ \textrm{meV}$ below $E_\mathrm{F}$ (see panels~\textbf{h,i}). The expected connection between the bulk-related features at $\Gamma$ and $M$ (\textbf{d}) is clearly seen and indicated with a dashed line and black arrows.~\textbf{f,g,} EMCs near the $M$ point. Dashed lines in~\textbf{f} show the prediction of {\it ab initio} calculations renormalized by a factor of 1.8 and shifted towards $E_\mathrm{F}$, see panel~\textbf{d} and ref.~\onlinecite{Charnukha_SFCAO_2015}. White line and arrows in~\textbf{g} indicate an additional electron band distinct from the complex structure in~\textbf{f} and producing the circular feature around~$M$ in~\textbf{a,b}.~\textbf{h,i,} EMCs near $\Gamma$ at $1~\textrm{K}$ and $50~\textrm{K}$, respectively. In~\textbf{h}, a hole band just below $E_\mathrm{F}$ (upper dashed line) and two more bands terminating at $56\ \textrm{meV}$ binding energy (lower dashed lines; see supplementary information for conclusive evidence for the two-band character of this feature) are indicated. Panel~\textbf{i} shows the normal-state dispersion of the upper bulk hole band extracted from the fit of momentum-distribution curves (white circles) and a parabolic fit to this dispersion (black solid line). The latter provides the location of the band edge at $2.3\ \textrm{meV}$ below $E_\mathrm{F}$. All dashed and solid lines are schematic unless stated otherwise.}
\end{figure*}
\begin{figure*}[!t]
\includegraphics[width=\textwidth]{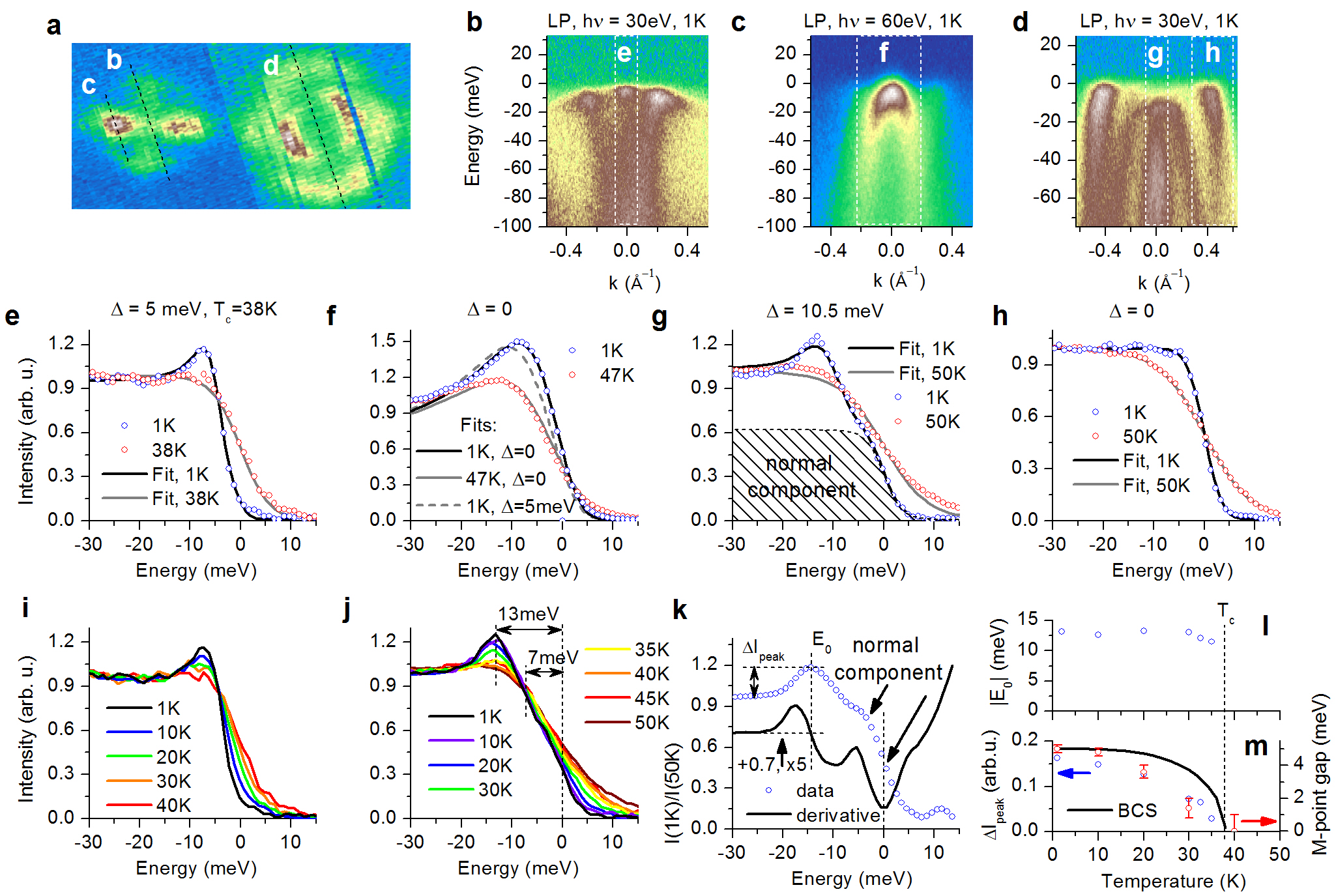}
\caption{\label{fig:sc}
~\textbf{a,} Crop of Fig.~\ref{fig:fs}a.~\textbf{b}--\textbf{d,} EMCs along the dashed lines in panel~\textbf{a}. All data were recorded at $1\ \textrm{K}$ using a linear polarization and $30\ \textrm{eV}$ (\textbf{b,d}) or $60\ \textrm{eV}$ (\textbf{c}) photon energy. Normal-state (at $50\ \textrm{K}$) counterpart of the EMC in~\textbf{d}, obtained under the same conditions, is shown in Fig.~\ref{fig:fs}i.~\textbf{e}--\textbf{h,} Normalized EDCs integrated in the momentum windows shown as dashed rectangles in panels~\textbf{b}--\textbf{d}. All fits were obtained using the Dynes--like model for integrated EDCs, as described in the text.~\textbf{i,j,} Temperature dependence of the normalized EDCs in~\textbf{e,g}. The dashed lines in~\textbf{j} indicate (left to right) the location of the quasiparticle coherence peak $E_0$, leading edge of the superconducting component at $1\ \textrm{K}$, and $E_\mathrm{F}$.~\textbf{k,} Ratio (blue open circles) of the normalized EDCs obtained in the superconducting ($1\ \textrm{K}$) and normal ($50\ \textrm{K}$) state in~\textbf{g}, and its first derivative (black solid line; shifted up by $0.7$ and multiplied by $5$ for clarity). Vertical dashed lines (left to right) mark the location of $E_0$ and $E_\mathrm{F}$. The former is further clearly visible as the zero of the first derivative (intersection of the left vertical and lower horizontal dashed lines). The normal component shown in panel~\textbf{g} produces an inflection point in the derivative and has its leading edge (minimum of the derivative) located at $E_\mathrm{F}$. It could originate in a finite contribution to the photoemission signal from outside the sample surface, given the small size of the single crystals (see Methods).~\textbf{l,m,} Temperature dependence of $\left|E_0\right|$ and $\Delta I_{\mathrm{peak}}$ defined in panel~\textbf{k} (blue circles) and of the $M$-point superconducting gap $\Delta$ (extracted from the data in~\textbf{b,e,i}) (red circles). Black solid line is the temperature dependence expected in the Bardeen-Cooper-Schrieffer theory of superconductivity for $\Delta=5\ \textrm{meV}$ and $T_{\mathrm{c}}=38\ \textrm{K}$. An accurate description of the temperature dependence of these superconducting features would require consistently taking into account the existence of multiple coupled bands and the proximity of their edges to the Fermi level.}
\end{figure*}

A typical photoemission intensity distribution at the Fermi level in \nfaofopt\ is shown in Figs.~\ref{fig:fs}a,b. It is dominated by two large circular sheets of the Fermi surface at the $\Gamma$ point of the Brillouin zone and a complex propellerlike structure surrounded by another circular sheet at the $M$ point. It has been shown in several works~\cite{Shen_LaFePO_ElStr_2008,HaiYun3761,PhysRevLett.101.147003,Liu2009491,Lu2009452,PhysRevLett.105.027001,PhysRevB.82.075135,PhysRevB.82.104519,Yang2011460,PhysRevB.84.014504}, particularly in ref.~\onlinecite{Charnukha_SFCAO_2015}, that all large circular features (two at the $\Gamma$ and one at the $M$ point) are extrinsic and can be traced back to the polar cleaved surface of the crystal. Below we provide additional compelling evidence that this is indeed the case. 

With the large circular features eliminated the electronic landscape that remains is clearly nontrivial. To examine it in detail, we show several energy-momentum cuts (EMCs) obtained along the dashed lines in Figs.~\ref{fig:fs}a,b. Figure~\ref{fig:fs}c reveals the photoemission intensity distribution along the high-symmetry $M$--$\Gamma$--$M$ direction. The surface-related bands that generate the large circular features in Figs.~\ref{fig:fs}a,b are shown with white arrows here and in subsequent panels of this figure. At the $\Gamma$ point, another hole band at higher binding energies is evident and is part of the bulk electronic structure, as will be substantiated in what follows. At the $M$ point, a very heavy hole band is clearly visible. This band gives rise to the blades of the propellerlike structure in Figs.~\ref{fig:fs}a,b, as has been shown in some of the earlier studies~\cite{PhysRevB.82.075135,Charnukha_SFCAO_2015}. Below, a weak electron band is visible with additional bands underneath it. In ref.~\onlinecite{Charnukha_SFCAO_2015} it has been argued that this entire intensity distribution is present already in the {\it ab initio} LDA calculations at much higher binding energies than found in experiment and is pulled to the Fermi level by electronic interactions, as is shown schematically in Fig.~\ref{fig:fs}d.

It is well-known that the photoemission matrix elements depend strongly on the polarization of incident light. Therefore, we show an energy-momentum cut along the same high-symmetry direction but obtained with the circular polarization in Fig.~\ref{fig:fs}e. Indeed, this figure reveals additional key features: the electron bands at the $M$ point are much more prominent and an additional intensity spot just below the Fermi level is visible, which comes from a higher-lying electron band, consistent with the band structure in Fig.~\ref{fig:fs}d. The photoemission intensity at the $\Gamma$ point now reveals a hole band just below the Fermi level as well as the additional intensity distribution with holelike dispersion at higher binding energies, already pointed out in Fig.~\ref{fig:fs}c. We show that the latter intensity distribution comprises two different bands degenerate at the $\Gamma$ point (see supplementary information), consistent with the theoretically predicted bulk electronic structure renormalized by a factor of about 1.8 as illustrated schematically in Fig.~\ref{fig:fs}d. One can clearly see that the upper band is connected to the heavy hole band at the $M$ point (as shown with black arrows and a dashed curve, displaced upwards not to conceal the weak photoemission signal). This observation is essential as it proves that the three hole bands at the $\Gamma$ point and the propeller structure at the $M$ point are part of the same electronic structure and are entirely distinct from the surface related bands mentioned above. Figure~\ref{fig:fs}f further shows that the band structure at the $M$ point agrees remarkably well with the prediction of {\it ab initio} calculations provided that the latter is renormalized by the same factor of 1.8 as the band structure around the $\Gamma$ point. It also indicates that albeit the hole bands giving rise to the blades of the propellerlike structure in the intensity maps in Figs.~\ref{fig:fs}a,b (integrated in a window of $\pm15\ \textrm{meV}$ around the Fermi level) terminate in the immediate vicinity of the Fermi level, they do not cross it and, therefore, do not contribute to the Fermi surface of \nfaofopt.

By taking an energy-momentum cut at $45^\circ$ with respect to the orientation of the blades, as shown in Fig.~\ref{fig:fs}g, we reveal the electron band that gives rise to the large circular feature at the $M$ point in Figs.~\ref{fig:fs}a,b. It can be seen that this electron band is much heavier than that shown in Fig.~\ref{fig:fs}f. In ref.~\onlinecite{Charnukha_SFCAO_2015} it has been shown that the electronic structure producing the propellerlike construct at the $M$ point does not have any intensity in this direction and thus this heavy electron band is not a part of it and must be extrinsic. Therefore, the only bulk band in \nfaofopt\ that does clearly cross the Fermi level and generates a well-defined Fermi surface is the shallow electron band at the $M$ point of the Brillouin zone, consistent with the predominantly electron, as opposed to hole, transport observed in this material~\cite{doi:10.1143/JPSJ.79.014710}. We thus arrive at the inherent electronic structure of \nfaofopt\ fully consistent with that shown in Fig.~\ref{fig:fs}d and surprisingly similar to that of \sfcaopt\ (ref.~\onlinecite{Charnukha_SFCAO_2015}). 

To determine the fine differences between these two compounds with very different superconducting transition temperatures we investigate the electronic structure more closely. Figure~\ref{fig:fs}h shows an energy-momentum cut at an angle to the $\Gamma$--$M$ direction, as indicated in Fig.~\ref{fig:fs}a. All the relevant features discussed above are now more immediately evident. The white arrows indicate the extrinsic surface states. The dashed lines, on the other hand, show the bulk hole bands identified above. One can clearly see the upper hole band in the immediate vicinity of the Fermi level crossing the inner surface-related band as well as the intensity distribution due to the lower two bands terminating at $\sim50\ \textrm{meV}$ below the Fermi level. The band edges of these three bands do not show any observable dispersion in the out-of-plane direction of the Brillouin zone (see supplementary information), confirming the highly two-dimensional character of the electronic structure of the 1111-type compounds predicted theoretically~\cite{PhysRevLett.100.237003,PhysRevB.81.155447}. To determine the precise location of the band edge closest to the Fermi level (and thus directly affecting the itinerant properties) we extract the dispersion of the corresponding band in the normal state (at $50\ \textrm{K}$) in Fig.~\ref{fig:fs}i from a fit of the momentum-distribution curves. A parabolic fit to the determined dispersion shows that this band terminates at only $2.3\ \textrm{meV}$ below the Fermi level and thus does not generate a Fermi surface.

Given that the overall bulk electronic structure of \nfaofopt\ is so strikingly similar to that of \sfcaopt, and yet their superconducting transition temperatures differ by a factor of 2, it is crucial to compare the effect of the fine differences on the superconducting pairing. To that end, we have studied the temperature dependence of the electronic structure around the $\Gamma$ and $M$ points of the Brillouin zone. Three representative energy-momentum cuts in the directions shown with dashed lines in Fig.~\ref{fig:sc}a are displayed in Figs.~\ref{fig:sc}b--d. In order to determine the superconducting energy gap accurately by accounting for the finite experimental resolution we use the expression derived in Ref.~\onlinecite{PhysRevB.79.054517}. The value of the superconducting energy gap is extracted by fitting the momentum-integrated EDCs in the momentum windows indicated in Figs.~\ref{fig:sc}b--d as white dashed rectangles with the Dynes function multiplied by the Fermi function and convolved with the resolution function:

\begin{equation}
\textrm{IEDC}(\omega)=\left[f(\omega,T)\left|Re\frac{\omega+i\Sigma^{\prime\prime}}{E}\right|\right]\otimes R_\omega(\delta E),\label{eq:dynes}\
\end{equation}
where $E=\sqrt{(\omega+i\Sigma^{\prime\prime})^2-\Delta^2}$, $\omega$ is the binding energy with reversed sign, $T$ is the temperature, $\Sigma^{\prime\prime}$ is the imaginary part of the self-energy, $\Delta$ is the superconducting energy gap, and $\delta E$ is the experimental resolution. The momentum resolution drops out from this expression due to momentum integration. The results of the fit are shown in Figs.~\ref{fig:sc}e-h for the middle and one of the blades of the propellerlike structure at the $M$ point, hole band terminating in the vicinity of the Fermi level at the $\Gamma$ point, as well as the outer surface-related band at the $\Gamma$ point, respectively. The superconducting energy gap at the center of the propellerlike structure amounts to $5\ \textrm{meV}$ and thus has a gap ratio $2\Delta/k_{\mathrm{B}}T_{\mathrm{c}}$ of about $3.2$, comparable to that on the corresponding feature of \sfcaopt\ ($\approx3.3$) (ref.~\onlinecite{Charnukha_SFCAO_2015}) and close to the BCS value $3.52$. The detailed temperature dependence of this superconducting gap, extracted by fitting the data in Fig.~\ref{fig:sc}i, is shown with red open circles in Fig.~\ref{fig:sc}m and appears to deviate from the behavior expected from the conventional Bardeen-Cooper-Schrieffer theory of superconductivity (black solid line). 

Unlike the energy-distribution curves recorded in the middle of the propellerlike structure, those obtained on one of the blades (Figs.~\ref{fig:sc}c and~f) exhibit pronounced finite-band effects, manifested in the strong deviation of the normal-state data in Fig.~\ref{fig:sc}f from the Fermi-Dirac distribution. In such a case Eq.~\ref{eq:dynes} cannot be used and the full modeling of a finite band must be employed~\cite{PhysRevB.79.054517}. The corresponding finite-band fits of the experimental data (open circles) in the normal and superconducting state (solid lines) are shown in Fig.~\ref{fig:sc}f and suggest the absence of the superconducting gap on this feature of the electronic structure, which, as has been argued above, does not produce any Fermi surface.

Quite surprisingly, the hole band terminating at $2.3\ \textrm{meV}$ {\it below} the Fermi level and thus likewise not producing any Fermi surface, is found to host a superconducting energy gap of $10.5\ \textrm{meV}$, with a gap ratio of $6.65$, significantly larger than the corresponding ratio in \sfcaopt\ ($\approx2.1$). Interestingly, this value is very close to the largest gap ratio of $\sim6.8$ observed in the optimally doped $\textrm{Ba}_{1-x}\textrm{K}_x\textrm{Fe}_2\textrm{As}_2$ compound with the same $T_{\mathrm{c}}\approx38\ \textrm{K}$ as \nfaofopt~\cite{0295-5075-83-4-47001,PhysRevB.79.054517}. Even though the extraction of the superconducting gap on this feature is somewhat complicated by the presence of a sizable non-superconducting (normal) component, as shown by the shaded area in Fig.~\ref{fig:sc}g and further substantiated in Fig.~\ref{fig:sc}k, the presence of a large superconducting gap is directly evident in the raw data presented in Figs.~\ref{fig:sc}d,g,j: the superconductivity-induced coherence peak is located at $13\ \textrm{meV}$ below the Fermi level and the leading edge of the superconducting component --- at about $7\ \textrm{meV}$ below the Fermi level (it is well-known that the leading-edge shift tends to underestimate the value of the superconducting gap). Both features vanish at the superconducting transition temperature, as can be seen in Fig.~\ref{fig:sc}j. The temperature dependence of the binding energy $E_0$ and amplitude of the coherence peak $\Delta I_{\mathrm{peak}}$, extracted from the experimental data and their derivative as indicated in Fig.~\ref{fig:sc}k, are presented in Figs.~\ref{fig:sc}l,m. It is clear that while $\Delta I_{\mathrm{peak}}$ (blue open circles in Fig.~\ref{fig:sc}m) closely mimics the temperature dependence of the superconducting gap found in the middle of the propellerlike structure at the $M$ point of the Brillouin zone (red open circles in Fig.~\ref{fig:sc}), the binding energy of the coherence peak shows negligible temperature dependence, suggesting filling-in, rather than closing, of the superconducting energy gap at the $\Gamma$ point.

Finally, and in stark contrast to previous studies, the outer surface-related band in our samples shows no sign of the superconducting energy gap, as shown in Fig.~\ref{fig:sc}h, providing further compelling evidence for the extrinsic character of this feature. The absence or strong suppression of the proximity effect between the bulk and the surface layer of \nfaofopt\ is remarkable, albeit not unprecedented~\cite{PhysRevLett.113.067003}, and requires further investigation.

The proximity of band-edge singularities to the Fermi level simultaneously at the $\Gamma$ and M points of the Brillouin zone, connected by the ubiquitous in the iron-based superconductors $(\pi,\pi)$ $Q$-vector, indicates that nesting may play an important role in the low-energy electrodynamics of this material. At the same time, the sizable value of the superconducting energy gap on the features of the electronic structure both at the $\Gamma$ and M points implies that the condition $\Delta/E_{\mathrm{F}}\ll1$, where $\Delta$ is the superconducting energy gap and $E_{\mathrm{F}}$ is the Fermi energy, required for the Migdal approximation of the conventional theories of superconductivity is violated. This suggests that a more general theory is necessary for an adequate description of the superconducting properties of this and likely other compounds of the iron-based family~\cite{Perali1996181,Bianconi_shaperesonance_2013}. 

Our observation of the dramatic enhancement of the superconducting energy gap in \nfaofopt\ given the band structure almost identical to that of \sfcaopt\ is striking. It leaves one with only two alternatives: either the superconducting pairing does not depend on the exact shape of the Fermi surface at all, or it must be extremely sensitive to very fine changes in the low-energy electronic structure of the iron oxypnictides. The former case is indeed possible and realized in conventional superconductors, in which the exact shape of the Fermi surface is largely irrelevant, given the same density of states at the Fermi level~\cite{Tinkham_superconductivity_1995_articlestyle}, and the superconducting pairing driven by a mediator not belonging to the electronic system in which superconductivity occurs, i.e. phonons. However, it has been shown early on that the sheer strength of the electron-phonon coupling is insufficient to explain the elevated superconducting transition temperatures in the iron oxypnictides~\cite{boeri:026403,PhysRevB.82.020506}. The density of states of two-dimensional parabolic bands (which is the case in \nfaofopt, as demonstrated in this work) is expected to be constant up to their respective band edges and thus would be very similar near the Fermi level in \nfaofopt\ and \sfcaopt. Therefore, one appears to be left with the latter option. If the pairing interaction originates in the same electronic subsystem that achieves the superconducting ground state then the simultaneous self-consistent account of the essential features of superconductivity and the experimentally observed highly singular electronic structure is indispensable.

\footnotesize

\mysection{Acknowledgements}
\noindent This project was supported by the German Science Foundation under Grants No. BO 1912/2-2 within SPP 1458. The work at PSI is supported by the Swiss National Science Foundation (No. 200021-137783) and the Sino-Swiss Science and Technology Cooperation (Project No. IZLCZ2138954). Work at ETH was partially supported by the SNSF and the NCCR program MaNEP. A.C. acknowledges financial support by the Alexander von Humboldt foundation.
\mysection{Author contributions}
\noindent A.C., D.V.E., C.E.M., N.X., and S.V.B. carried out the experiments. A.C., D.V.E., and S.V.B. analyzed the data. A.C. wrote the manuscript. N.D.Z. carried out the sample growth and characterization. S.V.B., M.S., B.Bg., and B.B. supervised the project. All authors discussed the results and reviewed the manuscript.
\mysection{Additional information}
\noindent {\bf Supplementary Information} accompanies this paper on www.nature.com/scientificreports.\\
\noindent {\bf Competing financial interests:} The authors declare no competing financial interests.\\
\noindent Correspondence should be addressed to A.C. (acharnukha@ucsd.edu) and S.V.B. (s.borisenko@ifw-dresden.de) Requests for materials should be addressed to N.D.Z. (zhigadlo@phys.ethz.ch) and B.Bg. (batlogg@solid.phys.ethz.ch).
\eject\newpage

\setcounter{figure}{0}
\renewcommand{\bibnumfmt}[1]{#1.}
\renewcommand{\bibnumfmt}[1]{S#1.}
\renewcommand\thefigure{S\arabic{figure}}
\renewcommand{\cite}[1]{[S\citenum{#1}]}

\thispagestyle{plainfancy}

\fontfamily{helvet}\fontseries{bf}\selectfont
\mathversion{bold}
\begin{widetext}
\begin{figure}
\vskip0pt\noindent\hskip-0pt
\hrule width\headwidth height\headrulewidth \vskip-\headrulewidth
\hbox{}\vspace{4pt}
\hbox{}\noindent\vskip10pt
\begin{center}
\huge\sffamily\textbf{Supplementary information}
\end{center}
\end{figure}
\end{widetext}

\normalfont\normalsize
\section{Materials and methods}
\par Angle-resolved photoemission measurements were performed at an angle-resolved photoemission spectroscope (``$1^3$--ARPES'' end-station at BESSY) operating below $900\ \textrm{mK}$~[S\onlinecite{Borisenko1cubedARPES_SRN2012},S\onlinecite{Borisenko1cubedARPESJove}], using synchrotron radiation within the range of photon
energies $20$-–$90\ \textrm{eV}$ and various polarizations on {\it in-situ} cleaved surfaces of high quality single crystals. The overall energy and angular resolution were $\sim 4\ \textrm{meV}$ and $0.3^\circ$, respectively, for the low-temperature measurements (FS mapping). The FS maps represent the momentum distribution of the intensity integrated within a $15\ \textrm{meV}$ window at the Fermi level. To equalize the intensity over different Brillouin zones the maps were normalised at each $\mathbf{k}$ point to the total recorded EDC intensity.
\par High-quality single crystals of superconducting \nfaofopt\ with masses of a few micrograms were synthesized by the high-pressure high-temperature cubic anvil technique and were characterized by x-ray diffraction, transport and magnetization measurements~\cite{PhysRevB.86.214509}. The latter revealed a superconducting transition temperature of about $38\ \textrm{K}$. The width of the superconducting transition was found to be less than $1\ \textrm{K}$, indicating very high homogeneity of the investigated samples.

\newpage
\section{Detailed low-energy electronic structure of \nfaofopt\ near the $\Gamma$ point}
\par
\begin{figure}[h]
\includegraphics[width=\textwidth]{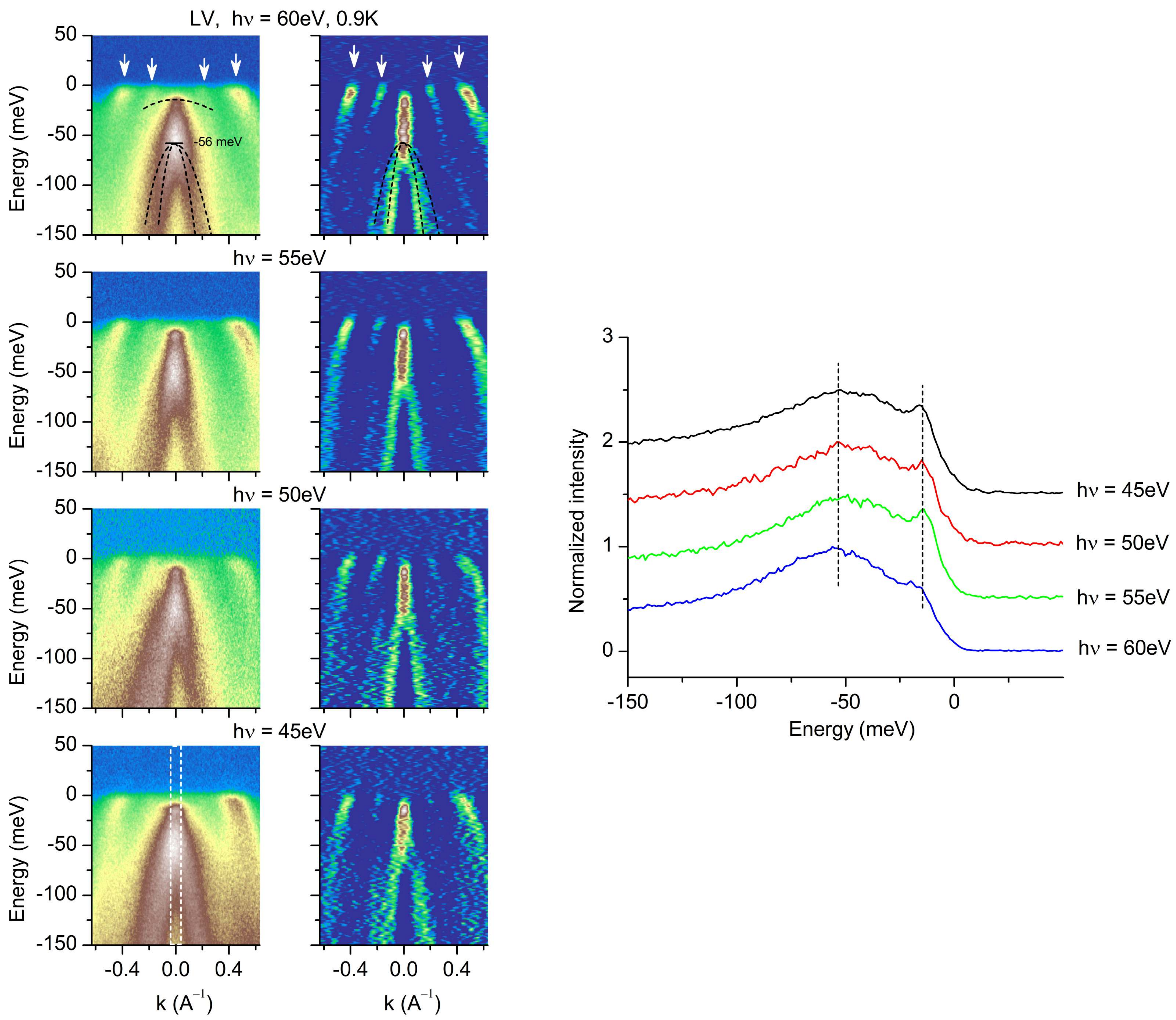}
\caption{\label{fig:kzdisp}
~\textbf{left column,} Energy-momentum cuts taken in the same direction as those in Fig.~1h,i using photons linearly polarized perpendicular to the plane of incidence and with decreasing energy from $60$ to $45\ \textrm{eV}$ in steps of $5\ \textrm{eV}$ (top to bottom). White arrows indicate the surface-related bands while the black dashed curves outline the three bulk hole bands, as discussed in the main text. Dashed rectangle in the bottom panel specifies the momentum window used to obtain the integrated energy-distribution curves shown in the right panel.~\textbf{middle column,} Second derivative of the data shown in the corresponding panels of the left column taken in the direction of the momentum. White arrows and dashed lines are the same as above.~\textbf{right panel,} Normalized integrated energy-distribution curves obtained by integration in the momentum window shown in the bottom panel of the left column.}
\end{figure}

 We have carried out measurements at various energies of the incident X-ray photons to investigate the dependence of the low-energy electronic structure of \nfaofopt\ on the out-of-plane momentum $\mathbf{k}_{\mathrm{z}}$. As shown in Fig.~\ref{fig:kzdisp}, neither the surface-related bands, nor, more importantly, the bulk hole band edges show any observable dependence on the photon energy and thus are strongly two-dimensional. Furthermore, the middle column of Fig.~\ref{fig:kzdisp} clearly demonstrates the presence of two separate bulk bands at binding energies below $56\ \textrm{meV}$, as argued in the main text and illustrated by dashed lines both in the present figure and in Fig.1 of the main text.
\par
\begin{figure}[h]
\includegraphics[width=\textwidth]{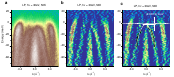}
\caption{\label{fig:gammader}
~\textbf{a,} Same data as in Fig.1i of the main text.~\textbf{b,c,} Second derivative of the experimental data in panel~\textbf{a}. The parabolic fit (white solid line) to the extracted quasiparticle dispersion of the top hole band at the $\Gamma$ point of the Brillouin zone (black open circles) is the same as in Fig.1i of the main text.}
\end{figure}

 Figure~\ref{fig:gammader} shows a magnified version of the data in Fig.1i of the main text and their derivative to demonstrate the quality of the Lorentz fit of the momentum-distribution curves (black circles in panel~\textbf{c}) as well as the parabolic fit to the thereby extracted quasiparticle dispersion of the top hole band at the $\Gamma$ point (white solid line in panel~\textbf{c}).
\newpage
\section{Temperature dependence of the photoemission intensity near the $\Gamma$ point}
\begin{figure}[h!]
\includegraphics[width=0.9\textwidth]{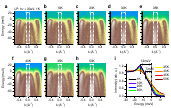}
\caption{\label{fig:gammaTdep}
~\textbf{a}--\textbf{h,} Temperature dependence of the photoemission intensity in the energy-momentum cut shown in Fig.2c. Dashed rectangle in all panels specifies the momentum window used to obtain the integrated energy-distribution curves shown in panel~\textbf{i}.~\textbf{i,} Energy-distribution curves integrated in the momentum window shown in panels~\textbf{a}--\textbf{h}. Vertical dashed lines indicate the position of (left to right) the superconducting coherence peak, the leading edge of the superconducting component, and the Fermi energy.}
\end{figure}
\newpage

\end{document}